\documentclass[aip,jcp,amsmath,amsymb,reprint]{revtex4-1}
\usepackage{graphicx}
\usepackage{bm}
\usepackage[utf8]{inputenc}
\usepackage[T1]{fontenc}
\usepackage{lmodern} 

\begin{document}

	\title{Breadth versus depth: Interactions that stabilize particle assemblies to changes in density or temperature}
	
	\author{William D. Piñeros}
	\affiliation{Department of Chemistry and Biochemistry, University of Texas at Austin, Austin, TX 78712}
	\author{Michael Baldea}
	\author{Thomas M. Truskett}
	\email{truskett@che.utexas.edu}
	\affiliation{McKetta Department of Chemical Engineering, University of Texas at Austin, Austin, TX 78712}
	\date{\today}
	\begin{abstract}
	    We use inverse methods of statistical mechanics to explore trade-offs associated with designing interactions to stabilize self-assembled structures against changes in density or temperature. Specifically, we find isotropic, convex-repulsive pair potentials that maximize the density range for which a two-dimensional square lattice is the stable ground state subject to a constraint on the chemical potential advantage it exhibits over competing structures (i.e., `depth' of the associated minimum on the chemical potential hypersurface). We formulate the design problem as a nonlinear program, which we solve numerically. This allows us to efficiently find optimized interactions for a wide range of possible chemical potential constraints.
	    We find that assemblies designed to exhibit a large chemical potential advantage at a specified density have a smaller overall range of densities for which they are stable. This trend can be understood by considering the separation-dependent features of the pair potential and its gradient required to enhance the stability of the target structure relative to competitors. Using molecular dynamics simulations, we further show that potentials designed with larger chemical potential advantages exhibit higher melting temperatures.
	\end{abstract}

	\maketitle 

\section{Introduction}
	The synthesis of matter with precise and well-characterized structures at nanometer length scales is critical to the discovery and manufacture of new material systems with desirable optical,\cite{QuantumDotPhotonic} mechanical\cite{NanoPiezoElectric}, and other physical\cite{InvDesignGeneral} properties. Unfortunately, despite recent progress,\cite{PhotonicMatsDesign, PhotonicMatsDesign2, PhotonicMatsDesign3} direct fabrication of such materials using top-down approaches can be technologically challenging, expensive, and time-consuming. An alternative bottom-up strategy is to synthesize systems of nanoscale particles with effective interactions\cite{ColloidInteractionsReview,SelfAssemblyPatchyParticles1} that favor spontaneous or directed self-assembly into the targeted structure (e.g., a periodic crystal or superlattice configuration).\cite{SelfAssemblySuperlatticeCatalysis,JanusParticlesSelfAssemblyRev,SelfAssemblyPolyhedraParticles,SelfAssemblySphericalColloidsPhotonic}

	Historically, design in self assembly has focused on {\em forward} approaches, whereby the interactions between particles are altered in some rationally-guided or Edisonian fashion, and the resulting equilibrium structures that such interactions produce are identified and catalogued according to their ability to meet various design goals concerning their material properties. However, attention has begun to shift toward {\em inverse} design strategies, whereby interactions that favor the targeted structure or properties in the thermodynamically stable state are formally discovered through a constrained, statistical mechanical optimization.\cite{InvDesignTechRev,InvDesignPerspective}

	Several recently introduced computational methods for inverse design focus on optimization of interparticle interactions to stabilize a targeted ground-state configuration with the assumption of an isotropic pair potential $\phi(r,\{\alpha_i\})$ with variable parameter set $\{\alpha_i\}$. Such approaches have found various interactions that stabilize two-dimensional square, honeycomb and kagome lattices\cite{MT_SquareHoneyConvexFull,RT_HoneyDoubleWell,InvDesignKagome,InvDesignKagomeFunctionalMethod} as well as the three-dimensional diamond crystal structure.\cite{MT_DiamondConvex, InvDesignKagomeDiamond} Furthermore, it was demonstrated that particles with the optimized interactions indeed assembled into the targeted lattice phases at higher temperature using molecular simulations \cite{MT_SquareHoneyConvexFull, RT_HoneyDoubleWell, InvDesignKagomeDiamond}.

In this same vein, we have used inverse methods to design convex-repulsive, isotropic pair potentials of the following form\cite{Avni3DLattices,AvniDimTransfer}
\begin{equation}
	\begin{aligned}
		\phi(r/\sigma) = \epsilon \lbrace A(r/\sigma)^{-n} &+ \sum_{i=1}^{2} \lambda_i(1-\tanh[k_i(r/\sigma-\delta_i)]) \\ 
								   &+f_{\text{shift}}(r/\sigma) \rbrace H[(r_{\text{cut}}-r)/\sigma].
	\end{aligned} 
\label{eq:potform}
\end{equation}
that stabilize open crystal structures. The final optimized potentials qualitatively resembled effective pair interactions observed in various soft colloidal systems (e.g., star polymers, ligand-passivated nanocrystals, microgels, etc.).\cite{ColloidInteractionsReview} Here, $\sigma$ and $\epsilon$ represent characteristic length and energy scales respectively; $H$ is the Heaviside function; $\{A,n,\lambda_i, k_i,\delta_i\}$ are variable parameters (i.e., ${\alpha_i}$), one of which is fixed to ensure $\phi(1)=\epsilon$; $r_{\text{cut}}$ is a cut-off radius; $f_{\text{shift}}$ is a quadratic function $f_{\text{shift}}(r/\sigma)= P (r/\sigma)^2 + Q r/\sigma + R$ added to enforce $\phi(r_{\text{cut}}/\sigma)= \phi'(r_{\text{cut}}/\sigma)= \phi''(r_{\text{cut}}/\sigma)= 0$. Using a simulated annealing optimization approach, parameters for this potential that stabilize, over a very wide range of density, square and honeycomb lattices in two dimensions\cite{AvniDimTransfer} as well as
simple cubic and diamond ground-state structures in three dimensions were determined.\cite{Avni3DLattices} Complete phase diagrams were also calculated for the three-dimensional systems,\cite{AvniPhaseDiagram3D} which illustrated rich and complex phase equilibria with the targeted assemblies exhibiting good thermal stability relative to other competing crystal lattices.

	One basic question that has not yet been addressed is, which features of a pair potential would tend to maximize the melting temperature of a given target structure? Moreover, how would encoding high thermal stability into the interactions affect the corresponding range of densities for which the target structure is favored? In other words, is there a natural compromise between designing for robustness to changes in temperature versus volume?  Such questions are challenging to answer directly via inverse methods because they would require incorporating full molecular simulations (for a wide range of model parameters and thermodynamic conditions) into the optimization problem, which is computationally unfeasible at present. A pragmatic alternative is to search for features of the ground-state behavior that, while easier to compute than higher temperature properties, correlate with thermal stability of the target phase.  In the present work, we find that placing constraints on the minimum chemical potential advantage that the target structure would exhibit over selected equi-pressure competing lattices at zero temperature helps determine optimized interactions with higher target-phase melting temperatures.

	The specific structure that we target via ground-state inverse optimizations in this work is the two-dimensional square lattice ground state, which has attracted considerable theoretical interest in recent years,~\cite{AvniDimTransfer,RampPotJagla1,GenRampPotJagla,ScalaStanleySquareWaterAnomalies} and the class of pair potentials we consider are those described by eq.~\ref{eq:potform}. The stable equilibrium ground-state structure can be established by determining the global minimum of the potential energy $U$ at fixed density and zero temperature or the minimum of the chemical potential $\mu$ at fixed pressure and zero temperature (amongst other possibilities that follow from classical thermodynamics\cite{Shellbook}). Following other works,~\cite{Avni3DLattices,AvniDimTransfer,B903931G,ZT_dHOpt} we adopt the latter fixed pressure framework for our optimizations for convenience because any coexistence between the target structure and another lattice also requires equality of pressure between phases.  Through our optimizations, we test how the maximum achievable range of density for stability of the square-lattice ground state is affected by constraining the differences between its chemical potential at a prescribed state point and those of selected competing lattice structures at the same pressure. To do this systematically requires the solution of a series of optimization problems, each utilizing different constraints. Given the considerable computational expense of using stochastic optimizers (e.g., simulated annealing, genetic algorithms, etc.) for even a single optimization, we instead formulate each optimization problem of interest as a constrained mixed-integer nonlinear problem, and we solve it numerically using the General Algebraic Modeling System (GAMS)\cite{GamsSoftware2013}. We then explore the consequences of our imposed chemical potential advantage of the target ground state for the resulting thermal stability (i.e., melting temperature) of the resulting lattices.

	The paper is organized as follows. We first introduce the computational methods used for carrying out the pair-potential optimization and the melting point estimation. Next, we show the relationship between the density range of ground-state stability for the target lattice and the minimum chemical potential difference between the target and selected competitors. We conclude by discussing how these trends manifest in the resulting pair potentials and lattice melting temperatures.

\section{Methods}
In this section, we describe how we formulate and solve the inverse design problem of interest in this work: finding isotropic pair interactions that maximize the density range over which the targeted square lattice is the ground-state configuration given a constraint on its chemical potential advantage over selected competing lattices. We further detail the implementation of molecular dynamics (MD) simulation methods for estimating the melting point to characterize the corresponding thermal stability of the designed lattice structures.

	\subsection{Inverse Design of the Pair Potential}
		\subsubsection{Optimization Problem Formulation}
We formulate the design optimization problem following the general paradigm
	\begin{align*}
		\underset{\text{decision variables}}{\text{maximize}} & \hspace{1cm} f(\mathbf{x})\\ 
		\text{subject to} &  \hspace{1cm}  g_i(\mathbf{x})	
	\end{align*}
where $f(\mathbf{x})$ is an objective function and $g_i(\mathbf{x})$ are constraint equality or inequality equations with variables $\mathbf{x}$. The mathematical forms of $f(\mathbf{x})$ and $g_i(\mathbf{x})$ define the type of problem to be solved (e.g., if integer variables or non-linear functions are necessary, etc.).  For the inverse design calculations of interest here, the set of equations $g_i$ incorporate any desired constraints to be imposed on the interparticle pair potential and $f$ is formulated to ensure optimization of the desired thermodynamic property. To optimize for specific ground states, one needs to consider not only the target lattice $l_t$ of the design, but also other lattices $\{l\}$ that naturally compete with it for thermodynamic stability (the procedure to determine the pool of competing lattices is discussed separately in the next section).
Using eq.\eqref{eq:potform} as the model pair potential, we ultimately seek potential parameters $\{A,n,l_i, k_i, d_i\}$ (i.e. the decision variables) that maximize the density range for which the target lattice $l_t$ has a chemical potential lower than lattices in $\{l\}$ at the same pressure such that a minimum specified chemical potential advantage of the target structure over select competitors is obtained at a given state point.

Specifically, to incorporate the pair potential of eq.\eqref{eq:potform}, we introduce constraint equations that ensure the potential is appropriately normalized, repulsive, convex, and continuous (we implicitly nondimensionalize energies by $\epsilon$, lengths by $\sigma$ and omit parameter notation below for brevity). The normality condition is given by
\begin{equation}
		\phi(1)=1
	\label{eq:potnorm}
\end{equation}
and the other constraints are given by
\begin{subequations}
	\begin{align}
		\phi(\mathbf{r})	&> 0 \\
		-\phi'(\mathbf{r}) 	&> 0 \\
		\phi''(\mathbf{r}) 	&> 0
	\end{align}
	\label{eq:potconvex}
\end{subequations}
and
	\begin{subequations}
	\begin{align}
		 \phi(r_{\text{cut}})	&=0	\\ 	
		 \phi'(r_{\text{cut}})	&=0	\\
		 \phi''(r_{\text{cut}}) 	&=0
	\end{align}
	\label{eq:potcutoff}
	\end{subequations}
We set $r_{\text{cut}}=2.27183$ as motivated by previous work considering square lattices designed via this potential form.\cite{AvniDimTransfer} As a practical matter, $\mathbf{r}$ is discretized over a finite set of points in $(0,r_{\text{cut}})$; we use ten uniform points in $\phi(\mathbf{r})$, and 60 points distributed in a 1:6:5 ratio from ranges [0.2,0.8), [0.8,1.2], (1.2,$r_{\text{cut}}$) for $\phi''(\mathbf{r})$, which we find sufficient to enforce the constraints. It is not necessary to include the constraint on $\phi'(\mathbf{r})$ so long as the other constraints are fulfilled.

Next, we specify the equations describing the physics of the ground state. The first is for the internal energy per particle, which can be expressed
	\begin{equation}	
		U_l=\frac{1}{2}\sum_{i}^{r_{i,l} \leq r_{\text{cut}}}  n_{i,l} \phi(r_{i,l}(\rho_l))	
    	\label{eq:latticesum}
	\end{equation}
Here, $r_{i,l}(\rho_l)$ are the density-dependent coordination distances for each lattice $l$ and $n_{i,l}$ are the number of neighbors at those distances.\cite{LatticeTableRef}
The pressure $P_l$ of lattice $l$ is related to its density $\rho_l$ by the virial expression
	\begin{equation}
		P_l = -\frac{1}{4} \rho_l \sum_{i}^{r_{i,l} \leq r_{\text{cut}}} n_{i,l} r_{i,l}(\rho_l) \phi'(r_{i,l}(\rho_l))
	\label{eq:pressure}
	\end{equation}
For our purposes, the relevant density of a competing lattice $l$, $\rho_l$, is that which leads to equality of pressure with the target lattice $l_t$ of density $\rho_t$.  In other words $\rho_l (\rho_t)$ can be determined from knowledge of $\rho_t$ via the relation $P_l (\rho_l)=P_t(\rho_t)$, and thus from eq.~\ref{eq:pressure}, we have
\begin{equation}
	\begin{aligned}
	\rho_l \sum_{i}^{r_{i,l} \leq r_{\text{cut}}} n_{i,l} r_{i,l}(\rho_l) \phi'(r_{i,l}(\rho_l))= \\ 
	\rho_t \sum_{i}^{r_{i,t} \leq r_{\text{cut}}} n_{i,t} r_{i,t}(\rho_t) \phi'(r_{i,t}(\rho_t))
	\end{aligned}
	\label{eq:rhol}
\end{equation}
The chemical potential of a ground-state lattice $l$ is, in turn, given by
	\begin{equation}
		\mu_l = U_l (\rho_l) + P_l(\rho_l)/\rho_l
	\label{eq:chempot}
	\end{equation}

Lastly, an auxiliary equation is used
	\begin{equation}
		{r_0^2(\rho_l)} \leq {r_{\text{cut}}^2}
	\end{equation}
where $r_0(\rho_l)$ represents the nearest neighbor distance for competing lattices at density $\rho_l$. This helps tighten the optimization formulation by keeping density within a reasonable range.

An objective function $f(\mathbf{x})$ that fulfills our optimization goals must also be specified. We choose such an objective function to evaluate to a finite scalar value $f(\mathbf{x}) \rightarrow f$ and to be conducive to maximizing the range of densities $\Delta\rho_t=\rho_{t,f}-\rho_{t,i}$ for which the target lattice exhibits a chemical potential lower than that of the competing lattices. This is then defined as 
\begin{equation}
		f = \sum_{\rho_t}\prod_{l} H[\mu_l(\rho_l(\rho_t))-\mu_t(\rho_t)]
		\label{eq_objfunc}
\end{equation}
where the sum is over discretized target lattice densities (each spaced a distance $\delta$ apart), $\rho_l(\rho_t)$ is computed from eq.~\ref{eq:rhol}, and $H$ is again the Heaviside step function.

An additional constraint equation,
	\begin{equation}
		\min\{\mu_l(\rho_l(\rho_{t,o}))\}-\mu_t(\rho_{t,o}) \geq \Delta
	\label{eq:deltaeq}
	\end{equation}
is introduced to specify the \emph{minimum} acceptable chemical potential difference $\Delta$ between the target lattice at an intermediate density point $\rho_{t,o}$ and selected competing lattices at the same pressure. Here, we use $\rho_{t,o}=1.39$, which was found in an earlier study\cite{AvniDimTransfer} to be in the middle of the density range of stability for a square lattice designed for $\Delta=0$ and the same pair potential form. In practice, we have found that the post-optimization chemical potential difference between the target lattice and its closest selected competitor  $\Delta\mu \equiv \min\{\mu_l(\rho_l(\rho_{t,o}))\}-\mu_t(\rho_{t,o})$ is approximately equal to the constraint $\Delta$ in all cases.  

\subsubsection{Numerical Solution Strategy}

We implemented the optimization problem described above in GAMS\cite{GamsSoftware2013}, using the Basic Open-source Nonlinear Mixed INteger (BONMIN) \cite{belotti2009branching,BONMIN} solver with Interior Point OPTimizer (IPOPT)\cite{wachter2006implementation} as the non-linear sub-solver. This choice of solver permits us to use integer valued functions such as in eq.\eqref{eq_objfunc} (i.e. the Heaviside function) as well as the remaining non-linear functions present in the potential and system physics formulation. 

In practice, each optimization begins by inputting an initial guess for the pair potential parameter set that does not violate the constraints of eq.\eqref{eq:potnorm}-\eqref{eq:potcutoff} and specifying a narrow target lattice density range $[\rho_{t,i},\rho_{t,f}]$ containing $\rho_{t,o}$ to consider. 

If the maximum attainable value of $f$ is realized in the optimization (i.e the maximum number of feasible density points $\Delta\rho_t/\delta=(\rho_{t,f}-\rho_{t,i})/\delta$ is achieved), the boundaries of the density range are widened and the previously attained potential is used as the initial guess for a new optimization. This procedure is repeated until the optimization returns $f<\Delta \rho_t/\delta$, indicating that the maximum density range of stability for a given chemical potential constraint $\Delta$ was attained in the previous optimization. We carry out the optimizations described above for different values of $\Delta$ to explore how an imposed chemical potential advantage of the target lattice affects the maximum attainable $\Delta \rho_t$. As discussed in the results section, there is a maximum value of $\Delta$ above which a feasible solution does not seem to exist for any density range. While found values are not verifiably global due to the local nature of the optimizer, they are optimal to the best of our efforts.  

In addition to the explicit constraints described above, only pair potentials that result in mechanically stable target ground-state structures (as determined from phonon spectra analysis) were considered. Spontaneous assembly of particles interacting via the optimized potentials from the fluid state into the target structure upon temperature quenching was also verified at $\rho_{t,o}$ using Monte Carlo simulations (see supplemental material). 

		\subsubsection{Competing Lattice Determination}
    In our previous work on inverse design of targeted lattices,\cite{Avni3DLattices,AvniDimTransfer} we sought pair potentials that simply maximized the density range of stability of the desired structure ($\Delta=0$). For that type of optimization, it was necessary to choose a finite pool of competitive structures to compare with the target lattice, ideally those with the lowest values of chemical potential at the pressures of interest (which are not generally known in advance). We determined the composition of this competitive pool from an iterative procedure. An initial set of structures was selected (e.g., Bravais lattices plus a small number of non-Bravais lattices or tilings) based on intuition and knowledge obtained from earlier simulation studies on similar pair potentials. An optimization was then performed using the chosen competitive pool, followed by a forward calculation of the ground-state phase diagram with the optimized pair potential for densities in the targeted range. Any new structures that appeared were subsequently added to the previous competitive pool, and a second optimization with the updated list of competitive structures was completed. This process--updating the competitive pool and optimizing the pair potential considering the expanded list of possible lattices found in forward calculations--was repeated until no new competing lattice structures emerged.

	In the present study, we repeat similar optimizations but with the added requirement of a minimum chemical potential difference between the target and selected competitors. The hypothesis is that such a constraint will find potentials displaying enhanced thermal stability of the targeted phase. Note that one cannot enforce a fixed chemical potential difference between the target lattice and {\em all} possible competitors. To understand why, consider a representation of lattice structure defined by a set of primitive and basis vectors $\{v\}$. If $\{v\}$ can be modified continuously in some way (without adding or removing particles), e.g., by a set of suitable parameters $\{\Theta\}$, then $\{v(\{\Theta\})\}$ will then define a hyperspace of continuously connected lattices with the target structure $l_t$ representing a specific point in this space. For ground-state systems of a given pair potential at a specific pressure, state quantities such as $\mu$ depend on the lattice structure (i.e. $\mu(\{v\})$) such that $\mu$ itself can be represented as a hypersurface of continuously connected lattices ${v(\{\Theta\})}$. Thus, one can always find structures in the neighborhood of the target lattice on the hypersurface with chemical potentials arbitrarily close to that of the target.
	
	 Considering this, it is clear that one cannot enforce a nonzero minimum chemical potential difference $\Delta$ between the target and \emph{all} possible competitive structures. However, one can meaningfully constrain the $\mu$ hypersurface in the optimization by enforcing a minimum chemical potential difference $\Delta$ between the target and a chosen set of lattices $\{l_g\}$ that define `flag points' on the landscape. This helps achieve a standardized and well defined constraint depth that is feasible for the optimization. Indeed, a similar approach was introduced by Zhang et al for a related $\Delta\mu$ optimization and justified under similar premises.\cite{ZT_dHOpt}
	
    We provide an example for concreteness. The chosen target square lattice can be represented as a point in a larger Bravais subspace spanned by oblique primitive vectors $\{v_B(\{\Theta\})\}$ with $\{\Theta\}$ consisting of an aspect ratio $b/a$ and primitive vector angle $\theta$. Thus, the square lattice is represented by ${v(1,\pi/2)}$, while other Bravais lattices like the triangular lattice are given by $\{v(1,\pi/3)\}$ and so on.  The corresponding $\mu(\{v\})$ landscape of this Bravais subset is then a function of $(b/a, \theta)$ (i.e. $\mu(\{v(b/a,\theta)\}$). As such, the flag-point lattices we choose for enforcing the depth constraints here are the triangular lattice and a rectangular lattice which capture independent variations along the $\theta$ and $b/a$ directions in the neighborhood of the target (for an illustration of the resulting Bravais chemical potential landscape from one of the optimized potentials see figure 1S).\cite{GenericSupplement} Similar subspace arguments can be made to account for elongated triangular (ET) and snub-square (SS) non-Bravais lattices, where the former can be transformed into square by a row-shift and the latter by a rotation of a single tile around its next neighboring square tiles. Other tilings or non-Bravais lattices can, in principle, be important for target lattice stability, but we did not find others that were relevant in the present square-lattice design problem.

	Given the above considerations, our final competing pool consisted of lattices determined from the iterative forward procedure, some of which (those which
	naturally belonged to a subspace that continuously deformed into the square lattice) were also chosen as flag-point lattices. From the Bravais subspace, the final competing pool consisted of triangular,  rectangular (REC) $b/a=1.17$, and oblique (OBL) $b/a=1.1$, $\theta=1.09$ lattices, with triangular and REC also serving as flag points for the chemical potential constraint. Similarly, for the other relevant non-Bravais subspaces, the competitive pool included one SS lattice ($b/a=1.0$) and three ET lattices ($b/a=1.07$,  $b/a=1.20$, and $b/a=1.23$), with all but the last ET lattice serving as flag points for the chemical potential constraint.

	\subsection{Melting Temperature Estimations}
	\subsubsection{Z-method}
	The Z-method is a microcanonical molecular dynamics (MD) simulation strategy for estimating the melting point of a crystal that does not require free energy calculations.\cite{ZmethodOrg}  The approach is based on the idea that a crystal remains metastable upon raising the temperature until it reaches its superheating limit, where it is hypothesized to have the same internal energy as the liquid at the equilibrium melting temperature $T_m$. The name is due to the fact that the estimate comes from the Z-shaped (zig-zag) graph that one observes for the system in the temperature $T$ vs pressure $P$ plane as it transforms from the solid phase upon raising the energy at constant volume. It has been applied for a variety of model systems and has been repeatedly tested for both accuracy and variability.\cite{ZmethodLenJohns,ZmethodEvals,ZmethodAluminumApp} For our purposes here, where we seek only estimates of melting temperatures to compare the widely varying thermal stabilities of targeted assemblies designed under various constraints, the Z-method provides an adequate guide.

	Energy sweeps for the Z-method are carried out as follows. Initial particle positions are set in a perfect square lattice, and initial velocities are chosen from a random distribution and scaled to achieve a desired initial kinetic energy. For a series of progressively increasing energies, microcanonical MD simulation trajectories of $N=1024$ particles (and a periodic square cell of length $V^{1/2}$ chosen to set $\rho=N/V=1.39$) are initiated with a time step value of $0.001$. After an initial pre-equilibration period at each energy, averages of static quantities like temperature $T$ and pressure $P$ are taken every $1000$ time steps for at least $10^6$ steps. Near the transition region, averages for liquid and solid properties are taken separately with the phase being determined by the translational order parameter $\tau$
	\begin{equation}
		\tau(k) = \frac{1}{N} \sum_{i=1}^N \cos{ (\mathbf{k} \cdot \mathbf{r}_i) }
	\end{equation}
 Here, $\mathbf{r}_i$ denotes particle positions vectors and $\mathbf{k}$ is a reciprocal lattice vector. We chose $\mathbf{k}=\frac{2\pi}{l} (1,1)$ for this purpose, where $l$ denotes the lattice constant value at density $\rho$. We use $\tau \ge 0.5$ to indicate solid configurations and $\tau \le 0.1$ to denote liquid configurations. These assignments were additionally supported by monitoring the mean square displacement of the particles as a function of time.

 Reported estimates of $T_m$ are averages of the temperature of the superheating limit of the solid obtained from twelve independent energy sweeps.
	\subsubsection{Hysteresis method}
	As further corroboration of the estimates obtained from the Z-method described above, we also carry out melting point estimations by the hysteresis method. This method is based on analysis of superheating and supercooling processes in the framework of nucleation theory and validated through molecular dynamics simulations.\cite{HysteresisMethod0,HysteresisMethod1} The basic approach is to carry out a simple heating and cooling sweep of the system near the melting point to determine the temperature of superheating $T_+$  and supercooling $T_-$. The melting point $T_m$ is then estimated from
	\begin{equation}
		T_m=T_+ + T_- -\sqrt{T_+ T_-}
	\label{eq:Tmhysteresis}   
	\end{equation}
	As such, we carry out Monte Carlo simulations in the canonical ensemble for $N=400$ particles in a periodic square cell of length $V^{1/2}$ adjusted to fix density at $\rho=N/V=1.39$. Simulations are started from the crystal phase and heated until melting is achieved. The system is then cooled from the liquid back into the crystal. The $T_+$ and $T_-$ points are obtained from the resulting hysteresis loop in an energy vs temperature diagram.   

\section{Results and Discussion}
\begin{figure}
	\includegraphics[scale=0.26]{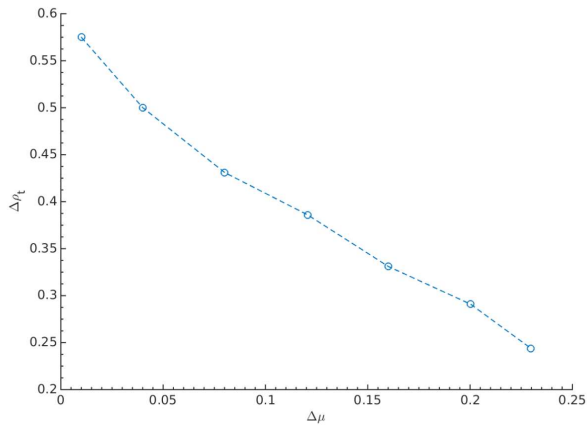}
	\caption{The width of the density range $\Delta\rho_t$ for which the square lattice is the stable ground-state structure for optimized parameters of the pair potential in eq.~\ref{eq:potform} versus the minimum chemical potential advantage $\Delta \mu$ of the square lattice ground state at $\rho_{t,o}$ over the flag-point lattices at that pressure. Blue circles indicate results using the solver BONMIN with IPOPT as the non-linear subsolver. Dashed lines are guides to the eye.}
	\label{fig:drhovsdmu}
	\end{figure}

	Using our described ground state optimization procedure, we were able to obtain pair potential parameters for eq.~\ref{eq:potform} that satisfied all of our objective goals. That is, we found potentials that a) were convex repulsive, b) maximized the density range $\Delta\rho_t$ for which the square lattice is the stable structure and c) were such that the target at density $\rho_{t,o}$ displayed a specified minimum chemical potential advantage  $\Delta\mu$ over the flag-point competitors (as elaborated in the methods section). The resulting relationship between $\Delta\rho_t$ subject to increasing values of $\Delta \mu$ for the optimized potentials is plotted in Figure~\ref{fig:drhovsdmu}.

     As seen, there is a clear negative correlation between $\Delta \mu$ and $\Delta \rho_t$. While the exact values of $\Delta\rho_t$ may change based on the choice of non-linear subsolver (also given local nature of the solutions), test runs using a different subsolver showed values that yielded a very similar trend (not shown). In other words, there appears to be a clear compromise with this potential form between designing for high stability at a given density and designing for stability with respect to changes in density. There also appears to be a limit with this potential form to how stable one can make the square lattice ground state at $\rho_{t,o}$ relative to the flag-point lattices ($\Delta \mu \approx 0.23$). For instance, we were only able to find solutions consistent with larger $\Delta\mu$ than those shown in Figure~\ref{fig:drhovsdmu} if we allowed the pair potential to violate the convexity constraint. 

     In terms of judging the overall quality of the optimizations, we can compare to one result from a previous study~\cite{AvniDimTransfer}, where a simulated annealing algorithm was used to find parameters for the potential of eq.~\ref{eq:potform} that maximized the range of densities for which the square lattice was the stable ground-state structure (with no chemical potential constraint). In that paper, $\Delta \rho_t=0.39$ was found for the optimized potential, which displayed a minimum chemical potential advantage of $\Delta\mu \approx 0.01$ over the flag-point lattices considered here. This can be compared to that of the potential obtained in this study  with a $\Delta\mu=0.01$ constraint, which exhibits a 50\% wider density range, $\Delta \rho_t=0.58$. While reported solutions are not verifiably global, the fact that such a large improvement in the objective function was obtained points to one of the advantages that the present rigorous framework has over heuristic optimization approaches like simulated annealing (a global optimizer in principle). 

	\begin{figure}
	\includegraphics[scale=0.26]{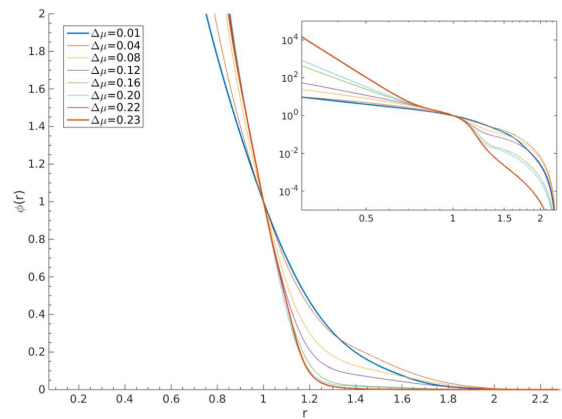}
	\caption{Optimized pair potentials $\phi(r)$ for different chemical potential constraints as a function of radial distance up to the cut-off at $r_{\text{cut}}$=2.27183. The inset shows a log-log plot of the same potentials.}
	\label{fig:potpars}
	\end{figure}
	 We now explore how features of the optimized interparticle potentials help to explain the observed trade-off associated with designing for a large chemical potential advantage of the target ground-state structure at a given density versus designing for target stability over a wide range of density. In Figure~\ref{fig:potpars}, the pair potentials corresponding to $\Delta\mu$=[0.01-0.23] are shown (for the full list of potential parameters values see tables S1 and S2)\cite{GenericSupplement}. While no pronounced features can be expected for strictly convex-repulsive interactions, two important aspects of the potential do manifest. As $\Delta \mu$ increases, so does the steepness of the core repulsion (for $r \lesssim 0.8$) as well as the rate of radial decay towards the cut-off point (for $r  \gtrsim 1.2$). The latter part can be seen more clearly in the log-log inset where intermediate features of core repulsion and radial decay can be seen to lie approximately between the two extrema potentials corresponding to $\Delta\mu=0.01$ and $\Delta\mu=0.23$. As we discuss next, this sharpening of radial-dependent features with increasing $\Delta\mu$ is what provides the chemical potential advantage of the target over its competitors, but at the cost of target lattice stability at other densities.

To look closer into the relation between pair potential form and target stability, it is helpful to recall that the chemical potential expression for a ground state system is given as $\mu_l = U_l + P_l/\rho_l$. Using the energy and pressure expressions in \eqref{eq:latticesum} and \eqref{eq:pressure}, it is possible to recast this expression as
	\begin{align}
	\mu_l &= \sum_{i}^{r_{i,l} \leq r_{\text{cut}}} n_{i,l} \left[ \frac{\phi(r_{i,l}(\rho_l))}{2}  - \frac{r_{i,l}(\rho_l)\phi'(r_{i,l}(\rho_l))}{4} \right] \\ \nonumber
	      &= \sum_{i}^{r_{i,l} \leq r_{\text{cut}}} n_{i,l}\psi(r_{i,l}(\rho_l))
    	\label{eq:chempot2}
	\end{align}
where $\psi(r)$ has been defined as
		\begin{equation}
			\psi(r) \equiv \frac{\phi(r)}{2}  - \frac{r \phi'(r)}{4}
			\label{eq:psir}
		\end{equation}
As such, we see chemical potential depends not only on the pair potential but also on its gradient. Analyzing the radial dependence of $\psi(r)$ will thus help to understand how the various lattice coordination shells at their respective radial separations contribute to the chemical potential and how they bias the functional form of the optimized potentials leading to the observed negative correlation between $\Delta\mu$ and $\Delta\rho_t$.
	\begin{figure*}
	\includegraphics[scale=0.26]{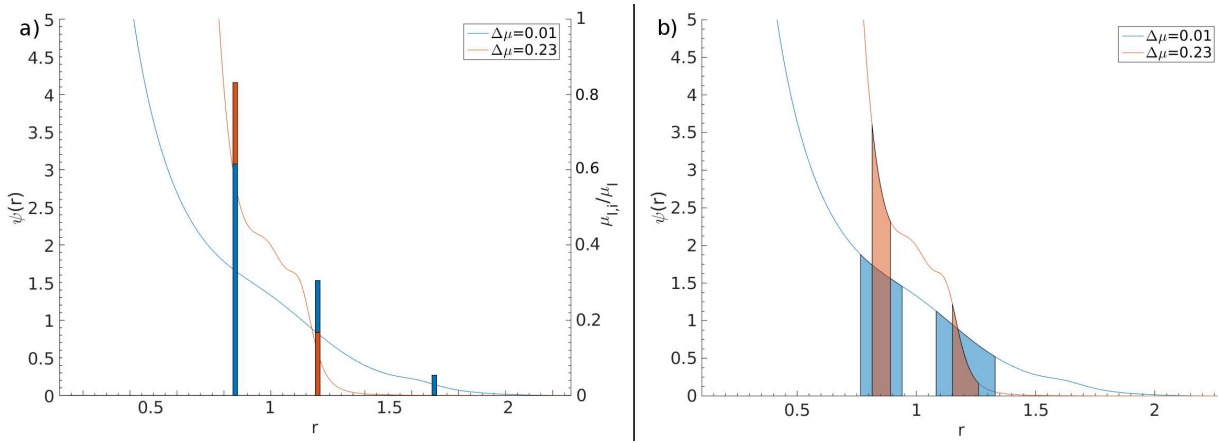}
	\caption{a) The function $\psi(r)$ of eq.~\ref{eq:psir} for optimized potentials with $\Delta\mu=0.01$ (blue) and $\Delta\mu=0.23$ (red), respectively. Bars indicate fractional contributions of each of the first three coordination shells to the total chemical potential for the square lattice at optimized density $\rho_o=1.39$. Bars are located at the respective coordination-shell distances. The contribution for the third coordination shell of the $\Delta\mu=0.23$ potential is not visible at this scale ($\sim10^{-4}$). 
		 b) $\psi(r)$ for the $\Delta\mu=0.01$ (blue) and $\Delta\mu=0.23$ (red) optimized pair potentials. Shaded areas indicate the ranges of the first and second neighbor distances (from left to right respectively) of the target lattice for densities where it is the stable ground-state structure.} 
	\label{fig:mu_neighbors_n_range}
	\end{figure*}

	We illustrate these points by plotting $\psi(r)$ for optimized interactions corresponding to the limiting cases of strongly ($\Delta\mu=0.23$) and weakly ($\Delta\mu=0.01$) constrained chemical potential advantage of the square lattice ground state over the flag point structures. The plot in figure \ref{fig:mu_neighbors_n_range}a) compares both $\psi(r)$ and the fractional coordination-shell contributions to the chemical potential of the square lattice for the two potentials. As can be seen, interactions obtained with the larger $\Delta\mu$ constraint impart greater emphasis on first-shell contributions that translate into potentials with harder cores and faster decays at these distances. These $\psi(r)$ features help the square lattice realize a lower chemical potential than the triangular lattice whose more densely packed first-coordination shell lies at a separation similar to that of the square lattice. Equally important is the shoulder-like region that decays between the square lattice's first and second coordination shells. The role of this shoulder is to destabilize the closely competitive rectangular and elongated triangular lattices that have second coordination shells at separations within the shoulder region and thus contribute to their higher values of chemical potential compared to that of the square lattice (See table S3 for a list of $\mu_{l,i}/\mu$ values of selected lattice competitors shells up to the third coordination)\cite{GenericSupplement}.  
	
	The potential shape trends obtained from optimizations with the high $\Delta\mu$ constraint described above can be contrasted to the muted features that manifest when a smaller $\Delta\mu$ constraint is applied (which leads to considerably larger $\Delta \rho_t$). Shown in figure \ref{fig:mu_neighbors_n_range}b) is also $\psi(r)$ for the two cases, but now presented along with shaded areas to indicate the range of first- and second-coordination shell distances of the corresponding stable square lattices. The key point is that small changes in coordination distances (due to changes in density) would have very different consequences for the chemical potential of the $\Delta \mu =0.23$ system as compared to the $\Delta \mu =0.01$ system due to their different forms for $\psi(r)$. For the $\Delta \mu =0.23$ system, small changes in density and coordination distances will produce pronounced changes in $\psi(r)$ and hence the chemical potential. As a result, the specific shape that provided great chemical potential advantage for the square lattice at $\rho_{t,o}$ is no longer able to favor the structure at even modestly lower or higher densities. In contrast, the slower varying form of $\psi(r)$ for the  $\Delta \mu =0.01$ system, while providing reduced chemical potential advantage at $\rho_{t,o}$, is able to keep the square lattice stable over a wider density range. The inverse relationship in figure \ref{fig:drhovsdmu} emerges as a natural consequence of this trade off.

	\begin{figure}
	\includegraphics[scale=0.26]{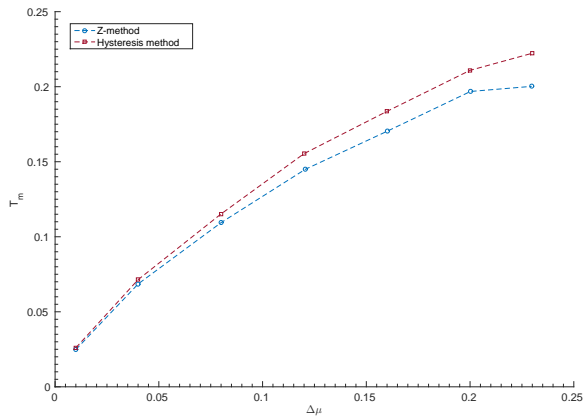}
	\caption{Estimated melting point of the targeted square lattice $\rho_{t,o}=1.39$ as a function of the minimum chemical potential advantage $\Delta \mu$ of the square lattice ground state at $\rho_{t,o}$ over the flag-point lattices at that pressure. Results obtained from the Z-method and the hysteresis method, respectively. Temperature in units of $\epsilon/k$. Dashed lines are guides to the eye.}
	\label{fig:Tmvsdmu}
	\end{figure}

    Moving on to understand how designing potentials for large $\Delta \mu$ for the square lattice ground state at $\rho_{t,o}=1.39$ affects the thermal stability of the target structure, we use the Z-method and the hysteresis method to estimate the corresponding melting temperatures $T_m$ at that density. As can be seen in Figure~\ref{fig:Tmvsdmu}, the potentials optimized with larger $\Delta\mu$ constraints also show higher $T_m$ irrespective of the estimation method. For instance, while the $\Delta\mu$=0.01 system has a melting point at around $T_m$=0.02, the melting point for the $\Delta\mu$=0.23 potential is approximately $T_m$=0.2--an order of magnitude greater. A slightly more pronounced but largely similar result is found from the hysteresis method ($T_m=0.02$ to $T_m=0.22$ at $\Delta\mu=0.01$ to $\Delta\mu=0.23$ respectively). Considering the $\psi(r)$ analysis presented above, this trend makes intuitive sense. Potentials designed with larger $\Delta \mu$ constraints impose greater penalties to target lattice deformation, and hence a higher average energy is required to move particles from their perfect lattice positions. This apparently translates directly to a higher melting point for the structure. A similar argument can be made based on the discussion of the $\mu$ hypersurface in the methods section. Since $\Delta\mu$ captures an effective `well depth' for the target structure, imposing higher $\Delta\mu$ has the effect of creating greater `restoring forces' on the target (i.e. higher eigenvalues of the $\mu$ Hessian).\cite{ZT_dHOpt} This results in increased mechanical stability at the ground state and a correspondingly higher melting point as shown here.

	Finally, an important question arises when comparing back to Figure \ref{fig:drhovsdmu}. Since we probed the melting points along a path where both `range' ($\Delta \rho_t$) and `depth' ($\Delta \mu$) change simultaneously, how does $T_m$ change if we hold a particular depth constant and vary the range, or vice versa? From our discussion so far, we expect that depth alone will determine the thermal trend while the range will be largely inconsequential. Indeed, test runs where we probed systems with the same depth but different ranges yielded scatter around a mean value, whereas holding range constant and varying depth produced melting points consistent with \ref{fig:Tmvsdmu} (not shown). Thus, for our inverse optimized pair potential, $\Delta\mu$ of the target in the ground state appears to strongly correlate with the thermal stability of the assembly, while the corresponding density range of stability has no such clear connection to the melting temperature.

\section{Conclusion}

    We have used inverse methods of statistical mechanics to gain new insights into the trade-off between designing interactions for stability of a target structure with respect to changes in temperature versus density.
    Specifically, we have explored the consequences of constraining the minimum chemical potential advantage of a target square lattice ground state at a prescribed density $\rho_{t,o}$ over select competitors ($\Delta\mu$ or `depth' on the $\mu$ landscape) while designing potentials that maximize the range of density where the target ground state is stable ($\Delta \rho_t$). The resulting constrained nonlinear optimization problem was solved numerically.  For the isotropic, convex-repulsive pair interactions considered here, pair potentials designed with a larger $\Delta \mu$ constraint exhibited a narrower range of density stability $\Delta \rho_t$. The reasons for this compromise are apparent when examining the radially-dependent forms of the optimized pair potentials and their gradients. To enable high stability at a given density, features in the potential and its derivative must align with specific coordination shells to help produce the desired differences in chemical potential. When such features are present, however, the resulting target structures can lose stability with even modest changes in density.

    We have also verified, via MD and Monte Carlo simulations, that potentials exhibiting ground states designed with larger $\Delta \mu$ constraints have higher melting temperatures at the target density. Preliminary tests further suggests that it is $\Delta \mu$ alone, and not $\Delta \rho_t$, that correlates with $T_m$.  Both results are in accord with the idea that $\Delta \mu$ constraints ensure restoring forces on the $\mu$ hypersurface that resist deformation (and ultimately melting) of the target structure.

\begin{acknowledgments}
T.M.T. acknowledges support of the Welch Foundation (F-1696) and the National Science
Foundation (CBET-1403768). We also gratefully acknowledge many insightful and productive conversations with Dr. Avni Jain about the ground-state calculations in this work, as well as her help with comparing our results to those obtained from earlier simulated annealing approaches to inverse design.
\end{acknowledgments}
		


\begin{thebibliography}{41}%
\makeatletter
\providecommand \@ifxundefined [1]{%
 \@ifx{#1\undefined}
}%
\providecommand \@ifnum [1]{%
 \ifnum #1\expandafter \@firstoftwo
 \else \expandafter \@secondoftwo
 \fi
}%
\providecommand \@ifx [1]{%
 \ifx #1\expandafter \@firstoftwo
 \else \expandafter \@secondoftwo
 \fi
}%
\providecommand \natexlab [1]{#1}%
\providecommand \enquote  [1]{``#1''}%
\providecommand \bibnamefont  [1]{#1}%
\providecommand \bibfnamefont [1]{#1}%
\providecommand \citenamefont [1]{#1}%
\providecommand \href@noop [0]{\@secondoftwo}%
\providecommand \href [0]{\begingroup \@sanitize@url \@href}%
\providecommand \@href[1]{\@@startlink{#1}\@@href}%
\providecommand \@@href[1]{\endgroup#1\@@endlink}%
\providecommand \@sanitize@url [0]{\catcode `\\12\catcode `\$12\catcode
  `\&12\catcode `\#12\catcode `\^12\catcode `\_12\catcode `\%12\relax}%
\providecommand \@@startlink[1]{}%
\providecommand \@@endlink[0]{}%
\providecommand \url  [0]{\begingroup\@sanitize@url \@url }%
\providecommand \@url [1]{\endgroup\@href {#1}{\urlprefix }}%
\providecommand \urlprefix  [0]{URL }%
\providecommand \Eprint [0]{\href }%
\providecommand \doibase [0]{http://dx.doi.org/}%
\providecommand \selectlanguage [0]{\@gobble}%
\providecommand \bibinfo  [0]{\@secondoftwo}%
\providecommand \bibfield  [0]{\@secondoftwo}%
\providecommand \translation [1]{[#1]}%
\providecommand \BibitemOpen [0]{}%
\providecommand \bibitemStop [0]{}%
\providecommand \bibitemNoStop [0]{.\EOS\space}%
\providecommand \EOS [0]{\spacefactor3000\relax}%
\providecommand \BibitemShut  [1]{\csname bibitem#1\endcsname}%
\let\auto@bib@innerbib\@empty
\bibitem [{\citenamefont {Luque}\ and\ \citenamefont
  {Marti}(2011)}]{QuantumDotPhotonic}%
  \BibitemOpen
  \bibfield  {author} {\bibinfo {author} {\bibfnamefont {A.}~\bibnamefont
  {Luque}}\ and\ \bibinfo {author} {\bibfnamefont {A.}~\bibnamefont {Marti}},\
  }\href@noop {} {\bibfield  {journal} {\bibinfo  {journal} {Nature Photonics}\
  }\textbf {\bibinfo {volume} {5}},\ \bibinfo {pages} {137} (\bibinfo {year}
  {2011})}\BibitemShut {NoStop}%
\bibitem [{\citenamefont {Armiento}\ \emph {et~al.}(2011)\citenamefont
  {Armiento}, \citenamefont {Kozinsky}, \citenamefont {Fornari},\ and\
  \citenamefont {Ceder}}]{NanoPiezoElectric}%
  \BibitemOpen
  \bibfield  {author} {\bibinfo {author} {\bibfnamefont {R.}~\bibnamefont
  {Armiento}}, \bibinfo {author} {\bibfnamefont {B.}~\bibnamefont {Kozinsky}},
  \bibinfo {author} {\bibfnamefont {M.}~\bibnamefont {Fornari}}, \ and\
  \bibinfo {author} {\bibfnamefont {G.}~\bibnamefont {Ceder}},\ }\href@noop {}
  {\bibfield  {journal} {\bibinfo  {journal} {Physical Review B}\ }\textbf
  {\bibinfo {volume} {84}},\ \bibinfo {pages} {014103} (\bibinfo {year}
  {2011})}\BibitemShut {NoStop}%
\bibitem [{\citenamefont {Mlinar}(2015)}]{InvDesignGeneral}%
  \BibitemOpen
  \bibfield  {author} {\bibinfo {author} {\bibfnamefont {V.}~\bibnamefont
  {Mlinar}},\ }\href@noop {} {\bibfield  {journal} {\bibinfo  {journal}
  {Annalen der Physik}\ }\textbf {\bibinfo {volume} {527}},\ \bibinfo {pages}
  {187} (\bibinfo {year} {2015})}\BibitemShut {NoStop}%
\bibitem [{\citenamefont {Soukoulis}\ and\ \citenamefont
  {Wegener}(2011)}]{PhotonicMatsDesign}%
  \BibitemOpen
  \bibfield  {author} {\bibinfo {author} {\bibfnamefont {C.~M.}\ \bibnamefont
  {Soukoulis}}\ and\ \bibinfo {author} {\bibfnamefont {M.}~\bibnamefont
  {Wegener}},\ }\href@noop {} {\bibfield  {journal} {\bibinfo  {journal}
  {Nature Photonics}\ }\textbf {\bibinfo {volume} {5}},\ \bibinfo {pages} {523}
  (\bibinfo {year} {2011})}\BibitemShut {NoStop}%
\bibitem [{\citenamefont {Corbitt}, \citenamefont {Francour},\ and\
  \citenamefont {Raeymaekers}(2015)}]{PhotonicMatsDesign2}%
  \BibitemOpen
  \bibfield  {author} {\bibinfo {author} {\bibfnamefont {S.~J.}\ \bibnamefont
  {Corbitt}}, \bibinfo {author} {\bibfnamefont {M.}~\bibnamefont {Francour}}, \
  and\ \bibinfo {author} {\bibfnamefont {B.}~\bibnamefont {Raeymaekers}},\
  }\href@noop {} {\bibfield  {journal} {\bibinfo  {journal} {Journal of
  Quantitative Spectroscopy \& Radiative Transfer}\ }\textbf {\bibinfo {volume}
  {158}},\ \bibinfo {pages} {3} (\bibinfo {year} {2015})}\BibitemShut {NoStop}%
\bibitem [{\citenamefont {Hossain}\ and\ \citenamefont
  {Gu}(2014)}]{PhotonicMatsDesign3}%
  \BibitemOpen
  \bibfield  {author} {\bibinfo {author} {\bibfnamefont {M.~M.}\ \bibnamefont
  {Hossain}}\ and\ \bibinfo {author} {\bibfnamefont {M.}~\bibnamefont {Gu}},\
  }\href@noop {} {\bibfield  {journal} {\bibinfo  {journal} {Laser Photonics
  Reviews}\ }\textbf {\bibinfo {volume} {8}},\ \bibinfo {pages} {233} (\bibinfo
  {year} {2014})}\BibitemShut {NoStop}%
\bibitem [{\citenamefont {Likos}(2001)}]{ColloidInteractionsReview}%
  \BibitemOpen
  \bibfield  {author} {\bibinfo {author} {\bibfnamefont {C.~N.}\ \bibnamefont
  {Likos}},\ }\href@noop {} {\bibfield  {journal} {\bibinfo  {journal} {Physics
  Reports}\ }\textbf {\bibinfo {volume} {348}},\ \bibinfo {pages} {267}
  (\bibinfo {year} {2001})}\BibitemShut {NoStop}%
\bibitem [{\citenamefont {Zhang}\ and\ \citenamefont
  {Glotzer}(2004)}]{SelfAssemblyPatchyParticles1}%
  \BibitemOpen
  \bibfield  {author} {\bibinfo {author} {\bibfnamefont {Z.}~\bibnamefont
  {Zhang}}\ and\ \bibinfo {author} {\bibfnamefont {S.~C.}\ \bibnamefont
  {Glotzer}},\ }\href@noop {} {\bibfield  {journal} {\bibinfo  {journal} {Nano
  Letters}\ }\textbf {\bibinfo {volume} {4}},\ \bibinfo {pages} {1407}
  (\bibinfo {year} {2004})}\BibitemShut {NoStop}%
\bibitem [{\citenamefont {Auyeung}\ \emph {et~al.}(2015)\citenamefont
  {Auyeung}, \citenamefont {Morris}, \citenamefont {Mondloch}, \citenamefont
  {Hupp}, \citenamefont {Farha},\ and\ \citenamefont
  {Mirkin}}]{SelfAssemblySuperlatticeCatalysis}%
  \BibitemOpen
  \bibfield  {author} {\bibinfo {author} {\bibfnamefont {E.}~\bibnamefont
  {Auyeung}}, \bibinfo {author} {\bibfnamefont {W.}~\bibnamefont {Morris}},
  \bibinfo {author} {\bibfnamefont {J.~E.}\ \bibnamefont {Mondloch}}, \bibinfo
  {author} {\bibfnamefont {J.~T.}\ \bibnamefont {Hupp}}, \bibinfo {author}
  {\bibfnamefont {O.~K.}\ \bibnamefont {Farha}}, \ and\ \bibinfo {author}
  {\bibfnamefont {C.~A.}\ \bibnamefont {Mirkin}},\ }\href@noop {} {\bibfield
  {journal} {\bibinfo  {journal} {Journal of the American Chemical Society}\
  }\textbf {\bibinfo {volume} {137}},\ \bibinfo {pages} {1658} (\bibinfo {year}
  {2015})}\BibitemShut {NoStop}%
\bibitem [{\citenamefont {Zhang}, \citenamefont {Luijten},\ and\ \citenamefont
  {Granick}(2015)}]{JanusParticlesSelfAssemblyRev}%
  \BibitemOpen
  \bibfield  {author} {\bibinfo {author} {\bibfnamefont {J.}~\bibnamefont
  {Zhang}}, \bibinfo {author} {\bibfnamefont {E.}~\bibnamefont {Luijten}}, \
  and\ \bibinfo {author} {\bibfnamefont {S.}~\bibnamefont {Granick}},\
  }\href@noop {} {\bibfield  {journal} {\bibinfo  {journal} {Annual Review of
  Physical Chemistry}\ }\textbf {\bibinfo {volume} {66}},\ \bibinfo {pages}
  {581} (\bibinfo {year} {2015})}\BibitemShut {NoStop}%
\bibitem [{\citenamefont {Damasceno}, \citenamefont {Engel},\ and\
  \citenamefont {Glotzer}(2012)}]{SelfAssemblyPolyhedraParticles}%
  \BibitemOpen
  \bibfield  {author} {\bibinfo {author} {\bibfnamefont {P.~F.}\ \bibnamefont
  {Damasceno}}, \bibinfo {author} {\bibfnamefont {M.}~\bibnamefont {Engel}}, \
  and\ \bibinfo {author} {\bibfnamefont {S.~C.}\ \bibnamefont {Glotzer}},\
  }\href@noop {} {\bibfield  {journal} {\bibinfo  {journal} {Science}\ }\textbf
  {\bibinfo {volume} {337}},\ \bibinfo {pages} {453} (\bibinfo {year}
  {2012})}\BibitemShut {NoStop}%
\bibitem [{\citenamefont {Zhao}\ \emph {et~al.}(2014)\citenamefont {Zhao},
  \citenamefont {Shang}, \citenamefont {Cheng},\ and\ \citenamefont
  {Gu}}]{SelfAssemblySphericalColloidsPhotonic}%
  \BibitemOpen
  \bibfield  {author} {\bibinfo {author} {\bibfnamefont {Y.}~\bibnamefont
  {Zhao}}, \bibinfo {author} {\bibfnamefont {L.}~\bibnamefont {Shang}},
  \bibinfo {author} {\bibfnamefont {Y.}~\bibnamefont {Cheng}}, \ and\ \bibinfo
  {author} {\bibfnamefont {Z.}~\bibnamefont {Gu}},\ }\href@noop {} {\bibfield
  {journal} {\bibinfo  {journal} {Accounts of Chemical Research}\ }\textbf
  {\bibinfo {volume} {47}},\ \bibinfo {pages} {3632} (\bibinfo {year}
  {2014})}\BibitemShut {NoStop}%
\bibitem [{\citenamefont {Torquato}(2009)}]{InvDesignTechRev}%
  \BibitemOpen
  \bibfield  {author} {\bibinfo {author} {\bibfnamefont {S.}~\bibnamefont
  {Torquato}},\ }\href@noop {} {\bibfield  {journal} {\bibinfo  {journal} {Soft
  Matter}\ }\textbf {\bibinfo {volume} {5}},\ \bibinfo {pages} {1157} (\bibinfo
  {year} {2009})}\BibitemShut {NoStop}%
\bibitem [{\citenamefont {Jain}, \citenamefont {Bollinger},\ and\ \citenamefont
  {Truskett}(2014)}]{InvDesignPerspective}%
  \BibitemOpen
  \bibfield  {author} {\bibinfo {author} {\bibfnamefont {A.}~\bibnamefont
  {Jain}}, \bibinfo {author} {\bibfnamefont {J.~A.}\ \bibnamefont {Bollinger}},
  \ and\ \bibinfo {author} {\bibfnamefont {T.~M.}\ \bibnamefont {Truskett}},\
  }\href@noop {} {\bibfield  {journal} {\bibinfo  {journal} {AlChE Journal}\
  }\textbf {\bibinfo {volume} {60}},\ \bibinfo {pages} {2732} (\bibinfo {year}
  {2014})}\BibitemShut {NoStop}%
\bibitem [{\citenamefont {Marcotte}, \citenamefont {Stillinger},\ and\
  \citenamefont {Torquato}(2011)}]{MT_SquareHoneyConvexFull}%
  \BibitemOpen
  \bibfield  {author} {\bibinfo {author} {\bibfnamefont {E.}~\bibnamefont
  {Marcotte}}, \bibinfo {author} {\bibfnamefont {F.}~\bibnamefont
  {Stillinger}}, \ and\ \bibinfo {author} {\bibfnamefont {S.}~\bibnamefont
  {Torquato}},\ }\href@noop {} {\bibfield  {journal} {\bibinfo  {journal}
  {Journal of Chemical Physics}\ }\textbf {\bibinfo {volume} {134}},\ \bibinfo
  {pages} {164105} (\bibinfo {year} {2011})}\BibitemShut {NoStop}%
\bibitem [{\citenamefont {Rechtsman}, \citenamefont {Stillinger},\ and\
  \citenamefont {Torquato}(2005)}]{RT_HoneyDoubleWell}%
  \BibitemOpen
  \bibfield  {author} {\bibinfo {author} {\bibfnamefont {M.~C.}\ \bibnamefont
  {Rechtsman}}, \bibinfo {author} {\bibfnamefont {F.~H.}\ \bibnamefont
  {Stillinger}}, \ and\ \bibinfo {author} {\bibfnamefont {S.}~\bibnamefont
  {Torquato}},\ }\href@noop {} {\bibfield  {journal} {\bibinfo  {journal}
  {Physical Review letters}\ }\textbf {\bibinfo {volume} {95}},\ \bibinfo
  {pages} {228301} (\bibinfo {year} {2005})}\BibitemShut {NoStop}%
\bibitem [{\citenamefont {Edlund}, \citenamefont {Lindgren},\ and\
  \citenamefont {Jacobi}(2011)}]{InvDesignKagome}%
  \BibitemOpen
  \bibfield  {author} {\bibinfo {author} {\bibfnamefont {E.}~\bibnamefont
  {Edlund}}, \bibinfo {author} {\bibfnamefont {O.}~\bibnamefont {Lindgren}}, \
  and\ \bibinfo {author} {\bibfnamefont {M.~N.}\ \bibnamefont {Jacobi}},\
  }\href@noop {} {\bibfield  {journal} {\bibinfo  {journal} {Physical Review
  Letters}\ }\textbf {\bibinfo {volume} {107}},\ \bibinfo {pages} {085503}
  (\bibinfo {year} {2011})}\BibitemShut {NoStop}%
\bibitem [{\citenamefont {Torikai}(2015)}]{InvDesignKagomeFunctionalMethod}%
  \BibitemOpen
  \bibfield  {author} {\bibinfo {author} {\bibfnamefont {M.}~\bibnamefont
  {Torikai}},\ }\href@noop {} {\bibfield  {journal} {\bibinfo  {journal}
  {Journal of Chemical Physics}\ }\textbf {\bibinfo {volume} {142}},\ \bibinfo
  {pages} {144102} (\bibinfo {year} {2015})}\BibitemShut {NoStop}%
\bibitem [{\citenamefont {Marcotte}, \citenamefont {Stillinger},\ and\
  \citenamefont {Torquato}(2013)}]{MT_DiamondConvex}%
  \BibitemOpen
  \bibfield  {author} {\bibinfo {author} {\bibfnamefont {E.}~\bibnamefont
  {Marcotte}}, \bibinfo {author} {\bibfnamefont {F.}~\bibnamefont
  {Stillinger}}, \ and\ \bibinfo {author} {\bibfnamefont {S.}~\bibnamefont
  {Torquato}},\ }\href@noop {} {\bibfield  {journal} {\bibinfo  {journal}
  {Journal of Chemical Physics}\ }\textbf {\bibinfo {volume} {138}},\ \bibinfo
  {pages} {061101} (\bibinfo {year} {2013})}\BibitemShut {NoStop}%
\bibitem [{\citenamefont {Edlund}, \citenamefont {Lindgren},\ and\
  \citenamefont {Jacobi}(2013)}]{InvDesignKagomeDiamond}%
  \BibitemOpen
  \bibfield  {author} {\bibinfo {author} {\bibfnamefont {E.}~\bibnamefont
  {Edlund}}, \bibinfo {author} {\bibfnamefont {O.}~\bibnamefont {Lindgren}}, \
  and\ \bibinfo {author} {\bibfnamefont {M.~N.}\ \bibnamefont {Jacobi}},\
  }\href@noop {} {\bibfield  {journal} {\bibinfo  {journal} {Journal of
  Chemical Physics}\ }\textbf {\bibinfo {volume} {139}},\ \bibinfo {pages}
  {024107} (\bibinfo {year} {2013})}\BibitemShut {NoStop}%
\bibitem [{\citenamefont {Jain}, \citenamefont {Errington},\ and\ \citenamefont
  {Truskett}(2013{\natexlab{a}})}]{Avni3DLattices}%
  \BibitemOpen
  \bibfield  {author} {\bibinfo {author} {\bibfnamefont {A.}~\bibnamefont
  {Jain}}, \bibinfo {author} {\bibfnamefont {J.~R.}\ \bibnamefont {Errington}},
  \ and\ \bibinfo {author} {\bibfnamefont {T.~M.}\ \bibnamefont {Truskett}},\
  }\href@noop {} {\bibfield  {journal} {\bibinfo  {journal} {Soft Matter}\
  }\textbf {\bibinfo {volume} {9}},\ \bibinfo {pages} {3866} (\bibinfo {year}
  {2013}{\natexlab{a}})}\BibitemShut {NoStop}%
\bibitem [{\citenamefont {Jain}, \citenamefont {Errington},\ and\ \citenamefont
  {Truskett}(2014)}]{AvniDimTransfer}%
  \BibitemOpen
  \bibfield  {author} {\bibinfo {author} {\bibfnamefont {A.}~\bibnamefont
  {Jain}}, \bibinfo {author} {\bibfnamefont {J.~R.}\ \bibnamefont {Errington}},
  \ and\ \bibinfo {author} {\bibfnamefont {T.~M.}\ \bibnamefont {Truskett}},\
  }\href@noop {} {\bibfield  {journal} {\bibinfo  {journal} {Physical Review
  X}\ }\textbf {\bibinfo {volume} {4}},\ \bibinfo {pages} {031049} (\bibinfo
  {year} {2014})}\BibitemShut {NoStop}%
\bibitem [{\citenamefont {Jain}, \citenamefont {Errington},\ and\ \citenamefont
  {Truskett}(2013{\natexlab{b}})}]{AvniPhaseDiagram3D}%
  \BibitemOpen
  \bibfield  {author} {\bibinfo {author} {\bibfnamefont {A.}~\bibnamefont
  {Jain}}, \bibinfo {author} {\bibfnamefont {J.~R.}\ \bibnamefont {Errington}},
  \ and\ \bibinfo {author} {\bibfnamefont {T.~M.}\ \bibnamefont {Truskett}},\
  }\href@noop {} {\bibfield  {journal} {\bibinfo  {journal} {Journal of
  Chemical Physics}\ }\textbf {\bibinfo {volume} {139}},\ \bibinfo {pages}
  {141102} (\bibinfo {year} {2013}{\natexlab{b}})}\BibitemShut {NoStop}%
\bibitem [{\citenamefont {Jagla}(1998)}]{RampPotJagla1}%
  \BibitemOpen
  \bibfield  {author} {\bibinfo {author} {\bibfnamefont {E.}~\bibnamefont
  {Jagla}},\ }\href@noop {} {\bibfield  {journal} {\bibinfo  {journal}
  {Physical Review E}\ }\textbf {\bibinfo {volume} {58}},\ \bibinfo {pages}
  {1478} (\bibinfo {year} {1998})}\BibitemShut {NoStop}%
\bibitem [{\citenamefont {Jagla}(1999)}]{GenRampPotJagla}%
  \BibitemOpen
  \bibfield  {author} {\bibinfo {author} {\bibfnamefont {E.}~\bibnamefont
  {Jagla}},\ }\href@noop {} {\bibfield  {journal} {\bibinfo  {journal} {Journal
  of Chemical Physics}\ }\textbf {\bibinfo {volume} {110}},\ \bibinfo {pages}
  {451} (\bibinfo {year} {1999})}\BibitemShut {NoStop}%
\bibitem [{\citenamefont {Scala}\ \emph {et~al.}(2000)\citenamefont {Scala},
  \citenamefont {Sadr-Lahijany}, \citenamefont {Giovambattista}, \citenamefont
  {Buldyrev},\ and\ \citenamefont
  {Stanley}}]{ScalaStanleySquareWaterAnomalies}%
  \BibitemOpen
  \bibfield  {author} {\bibinfo {author} {\bibfnamefont {A.}~\bibnamefont
  {Scala}}, \bibinfo {author} {\bibfnamefont {M.~R.}\ \bibnamefont
  {Sadr-Lahijany}}, \bibinfo {author} {\bibfnamefont {N.}~\bibnamefont
  {Giovambattista}}, \bibinfo {author} {\bibfnamefont {S.~V.}\ \bibnamefont
  {Buldyrev}}, \ and\ \bibinfo {author} {\bibfnamefont {H.}~\bibnamefont
  {Stanley}},\ }\href@noop {} {\bibfield  {journal} {\bibinfo  {journal}
  {Physical Review E}\ }\textbf {\bibinfo {volume} {63}},\ \bibinfo {pages}
  {041202} (\bibinfo {year} {2000})}\BibitemShut {NoStop}%
\bibitem [{\citenamefont {Shell}(2015)}]{Shellbook}%
  \BibitemOpen
  \bibfield  {author} {\bibinfo {author} {\bibfnamefont {M.~S.}\ \bibnamefont
  {Shell}},\ }\href@noop {} {\emph {\bibinfo {title} {Thermodynamics and
  Statistical Mechanics: An Integrated Approach}}}\ (\bibinfo  {publisher}
  {Cambridge University Press},\ \bibinfo {year} {2015})\ Chap.~\bibinfo
  {chapter} {11}\BibitemShut {NoStop}%
\bibitem [{\citenamefont {Prestipino}, \citenamefont {Saija},\ and\
  \citenamefont {Malescio}(2009)}]{B903931G}%
  \BibitemOpen
  \bibfield  {author} {\bibinfo {author} {\bibfnamefont {S.}~\bibnamefont
  {Prestipino}}, \bibinfo {author} {\bibfnamefont {F.}~\bibnamefont {Saija}}, \
  and\ \bibinfo {author} {\bibfnamefont {G.}~\bibnamefont {Malescio}},\ }\href
  {\doibase 10.1039/B903931G} {\bibfield  {journal} {\bibinfo  {journal} {Soft
  Matter}\ }\textbf {\bibinfo {volume} {5}},\ \bibinfo {pages} {2795} (\bibinfo
  {year} {2009})}\BibitemShut {NoStop}%
\bibitem [{\citenamefont {Zhang}, \citenamefont {Stillinger},\ and\
  \citenamefont {Torquato}(2013)}]{ZT_dHOpt}%
  \BibitemOpen
  \bibfield  {author} {\bibinfo {author} {\bibfnamefont {G.}~\bibnamefont
  {Zhang}}, \bibinfo {author} {\bibfnamefont {F.~H.}\ \bibnamefont
  {Stillinger}}, \ and\ \bibinfo {author} {\bibfnamefont {S.}~\bibnamefont
  {Torquato}},\ }\href@noop {} {\bibfield  {journal} {\bibinfo  {journal}
  {Physical Review E}\ }\textbf {\bibinfo {volume} {88}},\ \bibinfo {pages}
  {042309} (\bibinfo {year} {2013})}\BibitemShut {NoStop}%
\bibitem [{\citenamefont {{GAMS Development
  Corporation}}(2015)}]{GamsSoftware2013}%
  \BibitemOpen
  \bibfield  {author} {\bibinfo {author} {\bibnamefont {{GAMS Development
  Corporation}}},\ }\href {\url{http://www.gams.com/}} {\enquote {\bibinfo
  {title} {{General Algebraic Modeling System (GAMS) Release 24.2.3}},}\
  }\bibinfo {howpublished} {Washington, DC, USA} (\bibinfo {year}
  {2015})\BibitemShut {NoStop}%
\bibitem [{\citenamefont {Hirschfelder}, \citenamefont {Curtiss},\ and\
  \citenamefont {Bird}(1954)}]{LatticeTableRef}%
  \BibitemOpen
  \bibfield  {author} {\bibinfo {author} {\bibfnamefont {J.~O.}\ \bibnamefont
  {Hirschfelder}}, \bibinfo {author} {\bibfnamefont {C.~F.}\ \bibnamefont
  {Curtiss}}, \ and\ \bibinfo {author} {\bibfnamefont {R.~B.}\ \bibnamefont
  {Bird}},\ }\href@noop {} {\emph {\bibinfo {title} {Molecular Theory of Gases
  and Liquids}}}\ (\bibinfo  {publisher} {John Wiley \& Sons},\ \bibinfo {year}
  {1954})\BibitemShut {NoStop}%
\bibitem [{\citenamefont {Belotti}\ \emph {et~al.}(2009)\citenamefont
  {Belotti}, \citenamefont {Lee}, \citenamefont {Liberti}, \citenamefont
  {Margot},\ and\ \citenamefont {Waechter}}]{belotti2009branching}%
  \BibitemOpen
  \bibfield  {author} {\bibinfo {author} {\bibfnamefont {P.}~\bibnamefont
  {Belotti}}, \bibinfo {author} {\bibfnamefont {J.}~\bibnamefont {Lee}},
  \bibinfo {author} {\bibfnamefont {L.}~\bibnamefont {Liberti}}, \bibinfo
  {author} {\bibfnamefont {F.}~\bibnamefont {Margot}}, \ and\ \bibinfo {author}
  {\bibfnamefont {A.}~\bibnamefont {Waechter}},\ }\href@noop {} {\bibfield
  {journal} {\bibinfo  {journal} {Optimization Methods \& Software}\ }\textbf
  {\bibinfo {volume} {24}},\ \bibinfo {pages} {597} (\bibinfo {year}
  {2009})}\BibitemShut {NoStop}%
\bibitem [{BON()}]{BONMIN}%
  \BibitemOpen
  \href@noop {} {\enquote {\bibinfo {title} {{BONMIN}},}\ }\bibinfo
  {howpublished} {http://www.gams.com/dd/docs/solvers/bonmin.pdf, last
  retrieved 8/29/2014}\BibitemShut {NoStop}%
\bibitem [{\citenamefont {W{\"a}chter}\ and\ \citenamefont
  {Biegler}(2006)}]{wachter2006implementation}%
  \BibitemOpen
  \bibfield  {author} {\bibinfo {author} {\bibfnamefont {A.}~\bibnamefont
  {W{\"a}chter}}\ and\ \bibinfo {author} {\bibfnamefont {L.~T.}\ \bibnamefont
  {Biegler}},\ }\href@noop {} {\bibfield  {journal} {\bibinfo  {journal}
  {Mathematical programming}\ }\textbf {\bibinfo {volume} {106}},\ \bibinfo
  {pages} {25} (\bibinfo {year} {2006})}\BibitemShut {NoStop}%
\bibitem [{Gen()}]{GenericSupplement}%
  \BibitemOpen
  \href@noop {} {}\bibinfo {note} {{Available as supplementary
  material}}\BibitemShut {NoStop}%
\bibitem [{\citenamefont {Belonoshko}\ \emph {et~al.}(2006)\citenamefont
  {Belonoshko}, \citenamefont {Skorodumova}, \citenamefont {Rosengren},\ and\
  \citenamefont {Johansson}}]{ZmethodOrg}%
  \BibitemOpen
  \bibfield  {author} {\bibinfo {author} {\bibfnamefont {A.~B.}\ \bibnamefont
  {Belonoshko}}, \bibinfo {author} {\bibfnamefont {N.}~\bibnamefont
  {Skorodumova}}, \bibinfo {author} {\bibfnamefont {A.}~\bibnamefont
  {Rosengren}}, \ and\ \bibinfo {author} {\bibfnamefont {B.}~\bibnamefont
  {Johansson}},\ }\href@noop {} {\bibfield  {journal} {\bibinfo  {journal}
  {Physical Review E}\ }\textbf {\bibinfo {volume} {73}},\ \bibinfo {pages}
  {012201} (\bibinfo {year} {2006})}\BibitemShut {NoStop}%
\bibitem [{\citenamefont {Belonoshko}\ \emph {et~al.}(2007)\citenamefont
  {Belonoshko}, \citenamefont {Davis}, \citenamefont {Skorodumova},
  \citenamefont {Lundow}, \citenamefont {Rosengren},\ and\ \citenamefont
  {Johansson}}]{ZmethodLenJohns}%
  \BibitemOpen
  \bibfield  {author} {\bibinfo {author} {\bibfnamefont {A.~B.}\ \bibnamefont
  {Belonoshko}}, \bibinfo {author} {\bibfnamefont {S.}~\bibnamefont {Davis}},
  \bibinfo {author} {\bibfnamefont {N.}~\bibnamefont {Skorodumova}}, \bibinfo
  {author} {\bibfnamefont {P.}~\bibnamefont {Lundow}}, \bibinfo {author}
  {\bibfnamefont {A.}~\bibnamefont {Rosengren}}, \ and\ \bibinfo {author}
  {\bibfnamefont {B.}~\bibnamefont {Johansson}},\ }\href@noop {} {\bibfield
  {journal} {\bibinfo  {journal} {Physical Review B}\ }\textbf {\bibinfo
  {volume} {76}},\ \bibinfo {pages} {064121} (\bibinfo {year}
  {2007})}\BibitemShut {NoStop}%
\bibitem [{\citenamefont {Alfe}, \citenamefont {Cazorla},\ and\ \citenamefont
  {Gillan}(2011)}]{ZmethodEvals}%
  \BibitemOpen
  \bibfield  {author} {\bibinfo {author} {\bibfnamefont {D.}~\bibnamefont
  {Alfe}}, \bibinfo {author} {\bibfnamefont {C.}~\bibnamefont {Cazorla}}, \
  and\ \bibinfo {author} {\bibfnamefont {M.}~\bibnamefont {Gillan}},\
  }\href@noop {} {\bibfield  {journal} {\bibinfo  {journal} {Journal of
  Chemical Physics}\ }\textbf {\bibinfo {volume} {135}},\ \bibinfo {pages}
  {024102} (\bibinfo {year} {2011})}\BibitemShut {NoStop}%
\bibitem [{\citenamefont {Bouchet}\ \emph {et~al.}(2009)\citenamefont
  {Bouchet}, \citenamefont {Bottin}, \citenamefont {Jomard},\ and\
  \citenamefont {Zerah}}]{ZmethodAluminumApp}%
  \BibitemOpen
  \bibfield  {author} {\bibinfo {author} {\bibfnamefont {J.}~\bibnamefont
  {Bouchet}}, \bibinfo {author} {\bibfnamefont {F.}~\bibnamefont {Bottin}},
  \bibinfo {author} {\bibfnamefont {G.}~\bibnamefont {Jomard}}, \ and\ \bibinfo
  {author} {\bibfnamefont {G.}~\bibnamefont {Zerah}},\ }\href@noop {}
  {\bibfield  {journal} {\bibinfo  {journal} {Physical Review B}\ }\textbf
  {\bibinfo {volume} {80}},\ \bibinfo {pages} {094102} (\bibinfo {year}
  {2009})}\BibitemShut {NoStop}%
\bibitem [{\citenamefont {Luo}, \citenamefont {Strachan},\ and\ \citenamefont
  {Swift}(2004)}]{HysteresisMethod0}%
  \BibitemOpen
  \bibfield  {author} {\bibinfo {author} {\bibfnamefont {S.-N.}\ \bibnamefont
  {Luo}}, \bibinfo {author} {\bibfnamefont {A.}~\bibnamefont {Strachan}}, \
  and\ \bibinfo {author} {\bibfnamefont {D.~C.}\ \bibnamefont {Swift}},\
  }\href@noop {} {\bibfield  {journal} {\bibinfo  {journal} {Journal of
  Chemical Physics}\ }\textbf {\bibinfo {volume} {120}},\ \bibinfo {pages}
  {11640} (\bibinfo {year} {2004})}\BibitemShut {NoStop}%
\bibitem [{\citenamefont {Luo}\ \emph {et~al.}(2003)\citenamefont {Luo},
  \citenamefont {Ahrens}, \citenamefont {Cagin}, \citenamefont {Strachan},
  \citenamefont {Goddard},\ and\ \citenamefont {Swift}}]{HysteresisMethod1}%
  \BibitemOpen
  \bibfield  {author} {\bibinfo {author} {\bibfnamefont {S.-N.}\ \bibnamefont
  {Luo}}, \bibinfo {author} {\bibfnamefont {T.~J.}\ \bibnamefont {Ahrens}},
  \bibinfo {author} {\bibfnamefont {T.}~\bibnamefont {Cagin}}, \bibinfo
  {author} {\bibfnamefont {A.}~\bibnamefont {Strachan}}, \bibinfo {author}
  {\bibfnamefont {W.~A.}\ \bibnamefont {Goddard}}, \ and\ \bibinfo {author}
  {\bibfnamefont {D.~C.}\ \bibnamefont {Swift}},\ }\href@noop {} {\bibfield
  {journal} {\bibinfo  {journal} {Physical Review B}\ }\textbf {\bibinfo
  {volume} {68}},\ \bibinfo {pages} {134206} (\bibinfo {year}
  {2003})}\BibitemShut {NoStop}%
\end{thebibliography}
%

\end{document}